\documentclass[aps,prl,twocolumn,showpacs,showkeys,amsmath,amssymb,superscriptaddress,groupedaddress]{revtex4}

\usepackage{amssymb,amsmath}
\usepackage{graphicx}
\usepackage{epsfig}

\def\Tr{\mbox{Tr}\,}

\newcommand{\ee}{\end{equation}}
\newcommand{\be}{\begin{equation}}
\newcommand{\bea}{\begin{eqnarray}}
\newcommand{\eea}{\end{eqnarray}}
\newcommand{\eu}{{\rm e}}
\newcommand{\ii}{{\rm i}}

\newcommand{\Ord}{{\rm O}}
\newcommand{\sn}{{\rm sn}}
\newcommand{\cn}{{\rm cn}}
\newcommand{\dn}{{\rm dn}}

\def\XXint#1#2#3{{\setbox0=\hbox{$#1{#2#3}{\int}$}
     \vcenter{\hbox{$#2#3$}}\kern-.5\wd0}}

\begin{document}

\title{Essential singularity in the Renyi entanglement entropy of the one-dimensional $XYZ$ spin-$1\over2$ chain}

\author{Elisa Ercolessi}
\email{ercolessi@bo.infn.it}
\affiliation{Department of Physics, University of Bologna,
V. Irnerio 46, 40126 Bologna, Italy}
\affiliation{I.N.F.N., Sezione di Bologna, V. Irnerio 46, 40126 Bologna, Italy}

\author{Stefano Evangelisti}
\email{stefano.evangelisti@gmail.com}
\affiliation{Department of Physics, University of Bologna,
V. Irnerio 46, 40126 Bologna, Italy}
\affiliation{I.N.F.N., Sezione di Bologna, V. Irnerio 46, 40126 Bologna, Italy}

\author{Fabio Franchini}
\email{ffranchini@sissa.it}
\affiliation{SISSA, V. Bonomea 265, 34136 Trieste, Italy}
\affiliation{I.N.F.N., Sezione di Trieste, V. Bonomea 265, 34136 Trieste, Italy}

\author{Francesco Ravanini}
\email{ravanini@bo.infn.it}
\affiliation{Department of Physics, University of Bologna,
V. Irnerio 46, 40126 Bologna, Italy}
\affiliation{I.N.F.N., Sezione di Bologna, V. Irnerio 46, 40126 Bologna, Italy}

\date{\today}

\begin{abstract}
We study the Renyi entropy of the one-dimensional $XYZ$ spin-$1/2$ chain in the entirety of its phase diagram. The model has several quantum critical lines corresponding to rotated $XXZ$ chains in their paramagnetic phase, and four tricritical points where these phases join.
Two of these points are described by a conformal field theory and close to them the entropy scales as the logarithm of its mass gap.
The other two points are not conformal and the entropy has a peculiar singular behavior in their neighbors, characteristic of an {\it essential singularity}. At these nonconformal points the model undergoes a discontinuous transition, with a level crossing in the ground state and a quadratic excitation spectrum. We propose the entropy as an efficient tool to determine the discontinuous or continuous nature of a phase transition also in more complicated models.
\end{abstract}



\maketitle

In the past several years there has been a constantly increasing interest in quantifying and studying the entanglement of virtually every physical system \cite{amico2008, eisert2010, calabrese2009}.
This interest is not surprising, as entanglement provides valuable insights from many different perspectives.

As it is often measured as entanglement entropy, that is, the Von Neumann entropy of the reduced density matrix of a system ${\cal S} = - \Tr \hat{\rho} \ln \hat{\rho}$, its origin lies in the context of quantum information theory \cite{bennet1996, nielsenchuang}. Essentially, it quantifies the ``quantumness'' of a state and therefore it provides a measure of its suitability for efficient quantum algorithms, in the ongoing quest for quantum computing.

In the physics of strongly interacting systems, entanglement has been welcomed as a new interesting correlation function, different in nature compared to the traditional ones, due to its nonlocal structure. In particular, it has provided a new challenge for the integrable model community on one side \cite{peschel1999, jin2004, its2005} and has led to new, more efficient approaches for numerical simulations (tensor network states) on the other \cite{vidal2008, verstraete2008, gu2008}.
For statistical physics, the entanglement entropy has also been proposed as a numerically efficient way to characterize a system, due to its diverging behavior across a phase transition \cite{vidal2003, amico2006}.

If we concentrate on the bipartite entanglement, that is, the entanglement entropy between two complementary regions, a lot is understood about its qualitative behavior. In general, it satisfies the so-called {\it area law} \cite{eisert2010}, that is, it is proportional to the area of the boundary dividing the two regions. This is naively understood considering that the entanglement is carried by correlations, that, in general, decay exponentially with the distance with a rate given by the correlation length $\xi$. If the volume of the two regions is much bigger than $\xi$, entanglement is localized at the boundary. Hence the area law.
This behavior is modified when correlations decay more slowly, like close to phase transitions, where $\xi \to \infty$.

The one-dimensional (1D) case is most interesting. As the boundary between regions consists of individual points, in a massive phase the entropy saturates to a finite value for sufficiently large systems. For critical phases described by a conformal theory, correlations decay as power laws and the entropy diverges logarithmically with the size of the region \cite{calabrese2004}.  Close to conformal points, the behavior is still governed by a Conformal Field Theory (CFT) and the divergence is logarithmical in $\xi$ \cite{calabrese2010}.

In this letter, we argue that close to a nonconformal point of phase transition the entropy shows a dramatically different behavior, characteristic of an {\it essential singularity}. Thus, we propose the entropy as an efficient (analytical or numerical) tool to distinguish, for example, a first-order ferromagnetic Quantum Phase Transition (QPT) in a spin chain from higher-orders ones. This kind of ability can be very helpful in the study of a variety of other models with nonconformal critical (but still gapless) points, such as those described by spin-$1$ Hamiltonians \cite{degliesposti2003} or in fermionic systems \cite{franca2008} and many more.

An essential singularity for the entropy was already observed at the bicritical point of the $XY$ model in \cite{franchini2007}, where it was noticed that the conformal prediction fails because low-energy excitations have a quadratic dispersion relation. However, the $XY$ chain can be considered a toy model; peculiar behaviors like these can be simply a mathematical feature, that does not survive in more realistic systems. Here we will consider a fully interacting model that cannot be mapped into free systems, like the $XY$ chain and those of \cite{peschel2009}, that is, the $XYZ$ spin-$1/2$ chain, defined by the Hamiltonian
\begin{equation}
   \hat{H}_{XYZ} =
   -\sum_{n} \left( J_{x}\sigma_{n}^{x} \sigma_{n+1}^{x}
   + J_{y}\sigma_{n}^{y} \sigma_{n+1}^{y} + J_{z}\sigma_{n}^{z} \sigma_{n+1}^{z} \right) \; ,
   \label{eq:XYZ1}
\end{equation}
where $\sigma_{n}^{\alpha}$ ($\alpha=x,y,z$) are the Pauli matrices acting on the site $n$, the sum ranges over all sites $n$ of the chain and the constants $J_{x}$, $J_{y}$ and $J_{z}$ take into account
the degree of anisotropy of the model. Without any loss of generality, we can rescale the energy and set $J_x = 1$, $J_y = \Gamma$, and $J_z = \Delta$.
This model is the most general spin-$1/2$ 1D chain with nearest neighbor interactions and its rich phase diagram has both conformal and nonconformal critical points. Moreover, unlike \cite{franchini2007}, this is a truly interacting, although still integrable, model.

Following the standard procedure, we start from the ground-state wavefunction $\mid 0 \, \rangle$ of (\ref{eq:XYZ1}) and divide the system into two parts, which we take as the semi-infinite left and right chains, thus dividing the Hilbert space as $\mathcal{H}=\mathcal{H}_{R}\otimes\mathcal{H}_{L}$.
We introduce the reduced density matrix by tracing out one of the two half-chains, say $\mathcal{H}_{L}$:
\be
   \hat{\rho} \equiv \Tr_{\mathcal{H}_{L}} \mid 0 \, \rangle \langle \, 0 \mid \; .
\ee
As an entanglement estimator we will consider the Renyi entropy \cite{renyi1970}
\be
   {\cal S}_{\alpha} \equiv {1 \over 1-\alpha} \ln \Tr \hat{\rho}^\alpha \; ,
   \label{eq:renyi1}
\ee
which reduces to the Von Neumann entropy for $\alpha \to 1$ ($\alpha$ being just an additional parameter at our disposal).

In the thermodynamic limit, the reduced density matrix of the $XYZ$ model is known exactly, due to its connection with the Corner Transfer Matrices (CTM) of the zero-field 8-vertex model \cite{Baxter}. Using this connection, the entropy of the $XYZ$ chain was calculated in \cite{ercolessi2010}, where it was shown that $\hat{\rho}$ can be written in the form
\be
   \hat{\rho} = {1 \over {\cal Z}} \bigotimes_{j=1}^\infty
                \left( \begin{array}{cc}
                    1 & 0 \\
                    0 & x^j
                \end{array} \right) \; ,
   \label{XYZrho}
\ee
where $x \equiv \exp[-\pi\lambda/I(k)]$, $\mathcal{Z} \equiv \prod_{j=1}^{\infty} (1+x^{j})$ and $k$ and $\lambda$ are elliptic parametrizations of the coupling constants that, in the principal regime of the 8-vertex model, can be written as
\be
   \Gamma = \frac{1+k\;\sn^2 (\ii \lambda)}{1-k\;\sn^2 (\ii \lambda)} \; ,
   \quad
   \Delta = -\frac{\cn (\ii \lambda) \; \dn (\ii \lambda)}{1-k\;\sn^2 (\ii \lambda)} \; ,
   \label{eq:XYZ3bis}
\ee
with $0 \le k \le 1$, $0 \le \lambda \le I(k')$. $I(k)$ is the complete elliptic integral of the first kind and $k' = \sqrt{1 - k^2}$ is the conjugated elliptic parameter.

In \cite{peschel2009}, the entanglement was studied for systems akin to free particles, using the fact that $\hat{\rho}$ is of the form (\ref{XYZrho}) with eigenvalues $x_j = \eu^{-2 j \epsilon}$ or $x_j = \eu^{- (2 j+1) \epsilon}$, for the ordered or disordered phases, with $\epsilon$ characteristic of the model. From the CTM of \cite{Baxter} it is known that all integrable, local spin-$1/2$ chains have a reduced density matrix of the form (\ref{XYZrho}). Thus, the degeneracy of the eigenvalues of $\hat{\rho}$ is universal for integrable models (and given by a partitioning problem \cite{franchini2010}) and the entanglement is completely characterized by $\epsilon$. In the case of the $XYZ$ model, $\epsilon = \pi {\lambda \over 2 I(k)}$.

Further defining $q \equiv \eu^{-\epsilon}$, the Renyi entropy is
\be
   {\cal S}_{\alpha} =
   \frac{\alpha}{\alpha-1} \sum_{j=1}^{\infty} \ln \left ( 1 + q^{2 j} \right)
   + \frac{1}{1-\alpha} \sum_{j=1}^{\infty} \ln \left ( 1 + q^{2 j \alpha} \right) \; ,
   \label{eq:renyi4}
\ee
which we can rewrite in terms of elliptic $\theta$ functions \cite{franchini2008} as:
\bea
   {\cal S}_{\alpha} (\epsilon)  & = & \frac{\alpha}{6(1-\alpha)}
   \ln \frac{\theta_{4}(0,q)\theta_{3}(0,q)}{\theta_{2}^{2}(0,q)}
   \label{thetaS} \\
   && + \frac{1}{6(1-\alpha)}
   \ln \frac{\theta_{2}^{2}(0,q^\alpha)}{\theta_{3}(0,q^\alpha)\theta_{4}(0,q^\alpha)}
   - \frac{\ln(2)}{3} \; .
   \nonumber
\eea
This form, although completely equivalent to (\ref{eq:renyi4}), is most suitable for studying the behavior of the entropy in the phase diagram of the model, thanks to the vast literature on elliptic functions.

\begin{figure}
   \begin{centering}
       \includegraphics[width=\columnwidth]{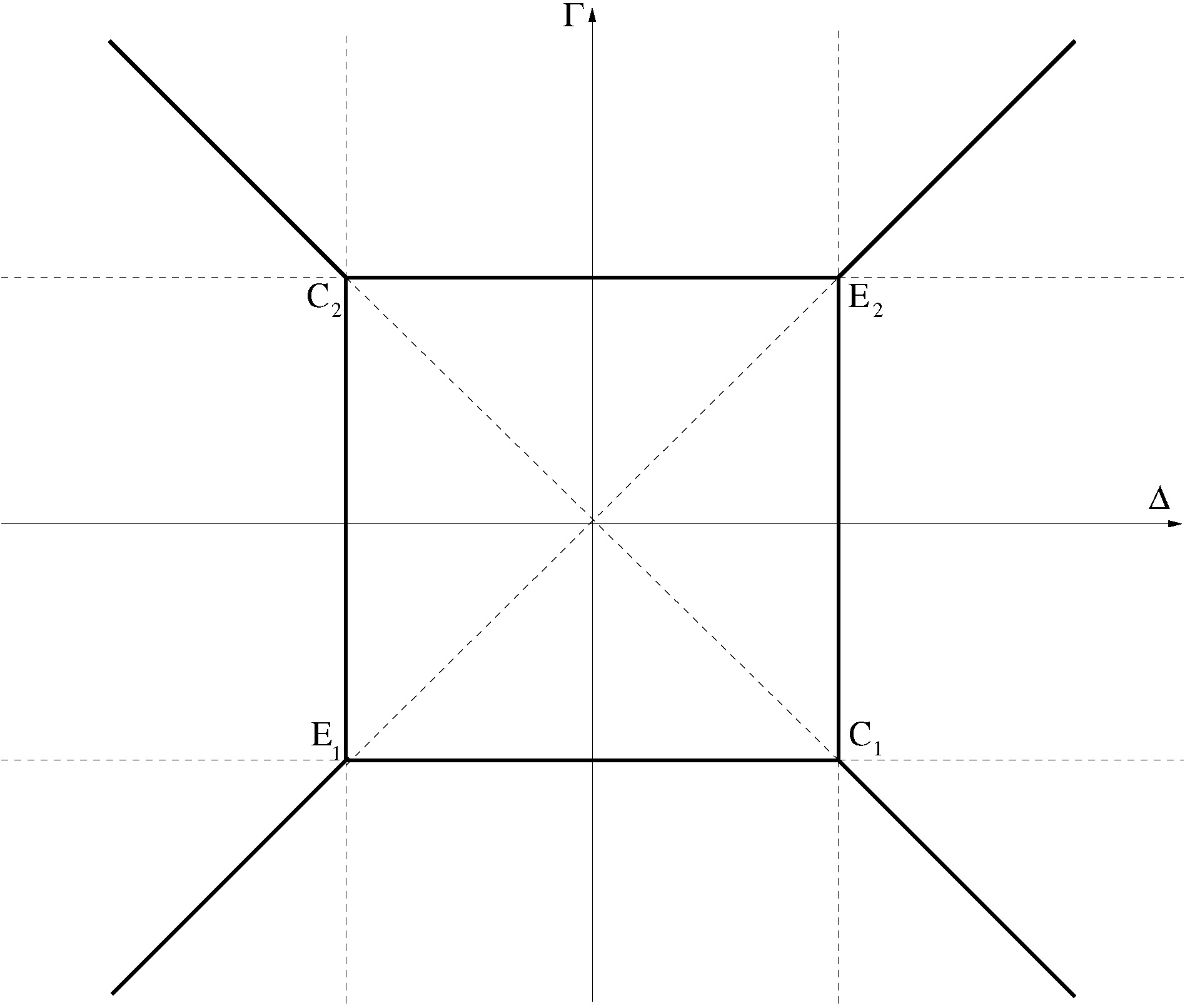}
   \end{centering}
   \caption{Phase Diagram of the XYZ model in the $(\Delta, \Gamma)$ plane. The solid lines --$\Delta = \pm 1$, $|\Gamma| \le 1$; $\Gamma= \pm 1$, $|\Delta| \le 1$ and $\Delta = \pm \Gamma$, $|\Delta| \ge 1$-- correspond to the critical phase of a rotated XXZ chain. Out of the four ``tricritical'' points, $C_{1,2}$ are conformal and $E_{1,2}$ are not.}
   \label{fig:phasediagram}
\end{figure}

In fig. \ref{fig:phasediagram} we plot a graph of the phase diagram of the $XYZ$ chain. The model is symmetric under reflections along the diagonals in the $(\Delta, \Gamma)$ plane. The system is gapped in the whole plane, except for six critical half-lines/segments: $\Delta = \pm 1$, $|\Gamma| \le 1$; $\Gamma= \pm 1$, $|\Delta| \le 1$ and $\Delta = \pm \Gamma$, $|\Delta| \ge 1$. All of these lines represent paramagnetic $XXZ$ chains, with the anisotropy along different directions. They are thus described by a $c=1$ (ultraviolet) sine-Gordon theory with $\beta$ dependent on $\Delta$ or $\Gamma$. These critical lines meet three by three at four points, that we will denote as ``tricritical'', since they separate three regions where the system is gapped. These three regions can be obtained one from the others by exchanging the role of the three spin components $\sigma^x$, $\sigma^y$, and $\sigma^z$.
Two of these points --$C_{1,2}=(1,-1),(-1,1)$-- are conformal points with $\beta^2 = 8 \pi$; while the other two --$E_{1,2}=(1,1),(-1,-1)$-- correspond to $\beta = 0$ and are {\it nonconformal}. The former points correspond to an antiferromagnetic Heisenberg chain at the BKT transition, while the latter correspond to a Heisenberg ferromagnet. Thus, at $E_{1,2}$ the system undergoes a transition in which the ground state passes from a disordered state to a fully aligned one. Exactly at the transition, the ground state is highly degenerate while the low-energy excitations are magnons with a quadratic dispersion relation.

\begin{figure}
   \begin{centering}
      \includegraphics[width=\columnwidth]{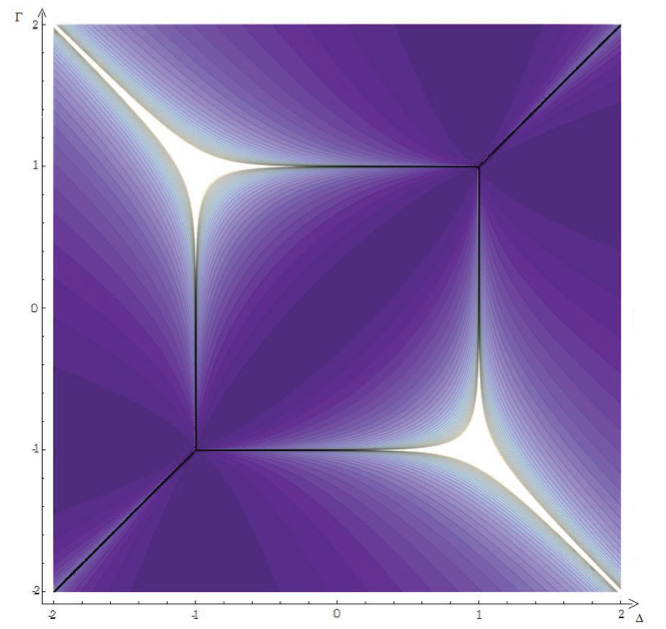}
   \end{centering}
   \caption{Contour plot of the Von Neumann entropy of the XYZ model in the $(\Delta,\Gamma)$ plane. Regions of similar colors have similar entropy values and the lines where colors change are lines of constant entropy. The brighter is the color, the bigger is the entropy.}
   \label{fig:isoentropy}
\end{figure}

The different behavior of the entropy close to these points is clear from fig. \ref{fig:isoentropy}, where we show a contour plot of the Von Neumann entropy as a function of $\Gamma$ and $\Delta$. In any neighbor of $C_{1,2}$ the entropy is diverging, since it is controlled by a conformal theory. Close to $E_{1,2}$ the entropy goes from $0$ to infinity depending on the direction of approach to the nonconformal point.
Moreover, we see that all curves of constant entropy pass through either $E_{1,2}$. This is the same behavior observed around the bicritical point of the $XY$ model in \cite{franchini2007}.

The nonconformal point of the latter model does not show a highly degenerate ground state, however, the density of states of low-energy excitations always diverges like $n(\varepsilon) \simeq 1 / \sqrt{\varepsilon}$ for a quadratic dispersion $\varepsilon(k) \simeq k^2$. Thus, in both models the ground state close to the multicritical point can be constructed out of a thermodynamically large number of configurations. This argument provides a qualitative picture of how this rich variety in the entanglement may arise close to nonconformal points.

For the $XYZ$ model we can make these statements more quantitative by expanding the parameters around the tricritical points.
Close to $C_2$ we take
\be
   \Gamma = 1 - \delta \cos \phi \; , \quad
   \Delta = -1 - \delta \sin \phi \; , \quad
   0 \le \phi \le {\pi \over 2} \; ,
   \nonumber
\ee
yielding
\bea
   k & = & \left( \sqrt{\tan \phi + 1} - \sqrt{\tan \phi} \right)^2 + {\rm O} \left( \delta \right) \; ,
   \label{vanishinglambda} \\
   \lambda & = &
   \sqrt{ {\delta \cos \phi \over 2}}
   \left( \sqrt{\tan \phi + 1} - \sqrt{\tan \phi} \right) + {\rm O} \left( \delta^{3/2} \right) \; .
   \nonumber
\eea
Thus, $k$ is not defined at the tricritical point, since its value as $\delta \to 0$ depends on the direction $\phi$ of approach (for $\phi=0$, $k=1$ and for $\phi=\pi/2$, $k=0$), while $\lambda \to 0$ as $\delta \to 0$, irrespective of $\phi$. Hence, $\epsilon \sim \sqrt{\delta}$ and in any neighborhood of the conformal point the entropy is diverging. In particular, on the direction $\phi=0$, since $I(1) \to \infty$, the entropy is divergent for any $\delta$ (as expected, since this is a critical line).

Expanding instead around $E_1$ as
\be
   \Gamma = - 1 + \delta \cos \phi \; , \quad
   \Delta = -1 - \delta \sin \phi \; , \quad
   0 \le \phi \le {\pi \over 2} \; ,
   \nonumber
\ee
we find
\bea
   k & = & \left( \sqrt{\tan \phi + 1} - \sqrt{\tan \phi} \right)^2 + {\rm O} \left( \delta \right) \; ,
   \label{singularlambda} \\
   \sn ( \ii \lambda ) & = &
   \ii \sqrt{ {2 \over \delta \cos \phi} }
   \left( \sqrt{\tan \phi + 1} - \sqrt{\tan \phi} \right) + {\rm O} ( \delta^{1/2} ) \; .
   \nonumber
\eea
Thus, $k$ is exactly as in the previous case, while $\lambda$ has a very singular behavior. A regular expansion of $\lambda$ in powers of $\delta$ is not readily available. However, one can see that $\lambda \sim I(k')$ close to $E_1$ and thus $\epsilon \sim I(k')/I(k)$, which varies from $0$ for $\phi = 0$ to infinity for $\phi = \pi/2$.

As we have already mentioned, the behavior of the entropy is completely determined by $\epsilon$, thus, the Renyi entropy has an essential singularity at $E_{1,2}$ and can take any positive real value arbitrarily close to it. This unusual behavior can be understood qualitatively considering that this tricritical point is a highly symmetric one with a highly degenerate ground state. Thus, as the parameters are changed to pass through $E_{1,2}$, the ground state undergoes a level crossing and the entanglement can change in a discontinuous way. This effect is most dramatic along one of the $XXZ$ lines, where the entropy goes suddenly from diverging in the paramagnetic phase to vanishing in the ferromagnetic one.

From the examples of the $XYZ$ chain and of the $XY$ model \cite{franchini2007}, we can infer that points of discontinuous phase transitions show a characteristic singular behavior of the entanglement entropy, which can be easily detected numerically and might be used as an efficient tool to distinguish, for example, between first- and higher-order phase transitions in more complicated models \cite{degliesposti2003, franca2008}. It is important to remark that ${\cal S}_\alpha$ and the correlation length show very different behaviors close to $E_{1,2}$, since, as it is expected for any phase transition, $\xi$ diverges as one approaches the critical lines or points, without showing the essential singularity characteristic of the entropy \cite{ercolessi2011}. Thus, close to the discontinuous points $E_{1,2}$ the entropy formula derived in \cite{calabrese2004} is no longer valid: indeed it applies in the vicinity of all conformal points of the critical lines (including the BKT points $C_{1,2}$), while it fails if we are close to nonconformal points, such as $E_{1,2}$, where the spectrum of excitations has a quadratic dispersion relation for small momenta.

The behavior of the Renyi entropy close to a conformal point has been recently investigated in \cite{calabrese2010}, where the Poisson summation formula was applied to (\ref{eq:renyi4}) to obtain its asymptotic behavior. Eq. (\ref{thetaS}) is equally suitable for such an expansion, using the different representations of the Jacobi theta functions \cite{wittaker}. In \cite{calabrese2010}, the Ising model and the $XXZ$ chain were considered. Using the $XYZ$ and the elliptic properties, we can easily extend their analysis. For instance, we can expand (\ref{thetaS}) in powers of $t \equiv \eu^{-\pi^2/\epsilon}$ for $\epsilon \to 0$ close to $C_{1,2}$:
\bea
   {\cal S}_\alpha (\epsilon) & = &
   \frac{\pi^{2}}{24} \left( 1 + \frac{1}{\alpha} \right) {1 \over \epsilon}
   - \frac{1}{2} \; \ln2
   \label{correction6} \\
   && + \frac{\alpha}{(1-\alpha)} \left[ t + \frac{t^{2}}{2} + \Ord (t^{3}) \right]
   \nonumber \\
   && - \frac{1}{(1-\alpha)} \left[ t^{1/\alpha} + \frac{t^{2/\alpha}}{2} + \Ord \left( t^{3/\alpha} \right) \right] \; .
   \nonumber
\eea

Because, close to these two tricritical points we have
\be
   \xi^{-1} \simeq 4 \eu^{-\pi^{2}/2\epsilon} = 4 t^{1/2} \; ,
\ee
we can write
\bea
   S_{\alpha} & = & \frac{1}{12} \left( 1 + \frac{1}{\alpha} \right) \ln(\xi)
   - {1 \over 6} \left( 2 - {1 \over \alpha} \right) \ln 2
   \label{correction6} \\
   && + \frac{\alpha}{(1-\alpha)} \left[ \frac{\xi^{-2}}{16}+\frac{\xi^{-4}}{512}
   + \Ord ( \xi^{-6} )\right]
   \nonumber \\
   && - \frac{1}{(1-\alpha)} \left[  (4\xi)^{-2/\alpha} + {1 \over 2}(4\xi)^{-4/\alpha}
   + \Ord \left( \xi^{-6/\alpha} \right) \right] \; .
   \nonumber
\eea
We recognize the familiar leading logarithmic term, a nonuniversal constant, and a series of corrections, in agreement with the results of \cite{calabrese2010}. For large $\alpha$ the leading correction is given by $\xi^{-\chi/\alpha}$ with $\chi = 2$. This implies that the operator  responsible for this correction has conformal dimensions $(\Delta,\overline{\Delta})=(1,1)$ and is therefore a quadratic combination of the $SU(2)_1$ Kac-Moody currents $J,\overline{J}$.

{\it Conclusions} We have studied the bipartite Renyi entropy of the 1D $XYZ$ spin chain in its phase diagram, using exact analytical expressions derived from the integrability of the model. The entropy diverges on the critical lines: close to conformal points the divergence is logarithmical in the correlation length with power-law corrections. At the nonconformal points the entropy has an essential singularity.  We argued that this may be a characteristic feature of discontinuous  phase transition points that could allow to easily numerically discriminate between first- and higher-order phase transitions. It would be interesting to test this behavior in other models.

Expression (\ref{thetaS}) for the Renyi entropy is general for integrable models of spin $1/2$ and possesses very interesting elliptic and modular properties, as a function of $k$, $\lambda$ and $\alpha$. \cite{ercolessi2011}
Different modular properties, in real space instead of parameter space, emerge close to the conformal points and have been used in \cite{calabrese2010} to understand the leading corrections to the entanglement entropy. It would be interesting to understand whether there is a link between these two.

\begin{acknowledgments}
{\it Acknowledgments} We wish to thank Marcello Dalmonte, Cristian Degli Esposti Boschi, Davide Fioravanti and Fabio Ortolani for useful and very pleasant discussions. We thank Ingo Peschel and Pasquale Calabrese for their helpful insights. F.F. would like to thank Prof. V. Korepin for several discussion prior to this work on the Essential Critical Point. This work was supported in part by two INFN COM4 grants (FI11 and NA41) and by the Italian Ministry of Education, University and Research grant PRIN-2007JHLPEZ.
\end{acknowledgments}

\vskip -0.5cm
%%%%%%%%%%%%%%%%%%%%%%%%%%%%%%%%%%%%%%

\end{document}